# Thermal stability of emission from single InGaAs/GaAs quantum dots at the telecom O-band

Paweł Holewa[1,*], Marek Burakowski[1,*], Anna Musiał[1, †], Nicole Srocka[2], David Quandt[2], André Strittmatter[2,**], Sven Rodt[2], Stephan Reitzenstein[2], and Grzegorz Sęk[1]

[1]Laboratory for Optical Spectroscopy of Nanostructures, Department of Experimental Physics, Wrocław University of Science and Technology, Wybrzeże Wyspiańskiego 27, 50-370 Wrocław, Poland

[2]Institute of Solid State Physics, Technische Universität Berlin, Hardenbergstraße 36, D-10623 Berlin, Germany

* These authors contributed equally

** Present address: Institute of Experimental Physics, Otto von Guericke University Magdeburg, D-39106 Magdeburg, Germany

[†] corresponding author: anna.musial@pwr.edu.pl

## Abstract

Single-photon sources are key building blocks in most of the emerging secure telecommunication and quantum information processing schemes. Semiconductor quantum dots (QD) have been proven to be the most prospective candidates. However, their practical use in fiber-based quantum communication depends heavily on the possibility of operation in the telecom bands and at temperatures not requiring extensive cryogenic systems. In this paper we present a temperature-dependent study on single QD emission and single-photon emission from metalorganic vapour-phase epitaxy-grown InGaAs/GaAs QDs emitting in the telecom O-band. Micro-photoluminescence studies reveal that trapped holes in the vicinity of a QD act as reservoir of carriers that can be exploited to enhance photoluminescence from trion states observed at elevated temperatures up to at least 80 K. The luminescence quenching is mainly related to the promotion of holes to higher states in the valence band and this aspect must be primarily addressed in order to further increase the thermal stability of emission. Photon autocorrelation measurements yield single-photon emission with a purity of $g^{(2)}_{50K}(0) = 0.13$ up to 50 K. Our results imply that these nanostructures are very promising candidates for single-photon sources at elevated temperatures in the telecom O-band and highlight means for improvements in their performance.

# Introduction

The field of quantum information science would greatly benefit from development of robust, efficient and on-demand single-photon sources (SPSs)[1,2]. In the past two decades, semiconductor quantum dots (QDs) have been investigated in that context in various material systems and have proven to be excellent single-photon emitters[3–5]. Epitaxial single In(Ga)As/GaAs QDs emitting below 1 μm have enabled the development of SPSs with record single-photon purity (i.e. with ultra-low probability of multiphoton emission events)[6,7], high indistinguishability[8–10], high photon extraction efficiency[11] and single-photon flux[12]. Most of these properties can also be achieved using electrical triggering[9,13] which brings the QD solution closer to real-world applications. QDs of this kind also benefit from compatibility with state-of-the-art GaAs-based semiconductor technology which provides mature and hence highly optimized growth, high structural quality, scalability and well developed methods for the fabrication of complex photonic structures (e.g., distributed Bragg reflectors – DBRs or photonic crystal cavities).

An important prerequisite for practical SPSs is compatibility with the existing telecommunication infrastructure based on standard silica fibres to enable low loss long-haul quantum communication in the relevant spectral range of 1.3 μm (O-band with local absorption minimum and vanishing dispersion) and 1.55 μm (C-band with global absorption minimum). One of the approaches is to utilize frequency conversion of near-infrared QD emission to shift it to 1.55 μm[14,15]. This is currently considered as the nearly optimal QD source and the down-conversion process can erase the small spectral difference between the two remote sources making photons originating from them indistinguishable, but the approach is rather complex and the conversion has still relatively low efficiency. In addition, it requires a laser of a very specific wavelength to be mixed with the QD emission inside a nonlinear medium. Therefore, it is appealing to develop sources directly emitting at the target wavelength. In the case of InAs QDs on GaAs the emission redshift to 1.3 μm is typically achieved via strain engineering and by, e.g., capping with a strain reducing layer (SRL)[16–21] or the nucleation on metamorphic buffer layers[22]. QDs with SRL have recently been used for the demonstration of a practical, compact, and fibre-integrated SPS at telecom O-band with working temperature of 40 K[23,24]. For SPSs emitting at 1.55 μm, the primarily considered candidates are single InAs/InP quantum dashes[25,26], InAs/InP QDs[27–30], and InAs/GaAs QDs based on the metamorphic layer approach[31,32]. Other approaches for telecom wavelength SPSs, despite their single-photon emission at room temperature, have their own drawbacks, e.g., atomic sources suffer from probabilistic nature of emission events[33], random direction of emission and low radiative rates. One has to keep in mind that widely used process of parametric down-conversion[34] is also a probabilistic photon generation scheme, which is also under development in the telecom.

The second pivotal requirement for practical, solid-state SPS for quantum communication is single-photon emission at significantly elevated temperatures of the emitter to make such quantum devices more compact, practical and cheaper. Reaching room temperature operation with QDs emitting at telecom wavelengths encounters fundamental limits of the quantum localization energy and inter-level spacing with the highest reported temperatures where the single-photon emission at telecom wavelengths is still observable at 120 K[35]. This is out of reach for thermoelectrical cooling, but on the contrary, the Stirling cryocooling can be successfully applied to reach temperatures as low as 27 K[23,24] and fulfils the requirements for compact, easy and relatively cheap operation without need for cryogenic liquids.

Studying the influence of temperature on the stability and purity of single-photon emission from QDs enables the determination of the highest working temperature for a given SPS design and the identification of processes responsible for quenching of QD emission and deterioration of single-photon purity. For single InGaAs/GaAs QDs with SRL emitting at 1.3 μm so far only Olbrich et al. carried out a temperature-dependent study[36]. They reported single-photon emission at liquid nitrogen temperature (77 K) with second order correlation function at zero time delay $g^{(2)}(0) = 0.21$ using the negative trion radiative recombination[36]. The structures investigated in this paper differ fundamentally from those investigated in Ref. [36]: In our case the emission is dominated by positively charged complexes due to unintentional carbon p-doping in the metalorganic chemical vapour deposition (MOCVD) process[37] in contrast to negative trions exploited in the case of previous study. Additionally, differences between the two group of nanostructures in QD morphology resulting from details of the growth procedure (i. e., exact growth parameters) are reflected in a modified electronic level structure and s-p energy splitting[38]. Moreover, in the present work the QDs are embedded in deterministically fabricated mesas which makes them relevant for high-yield single-photon sources and is crucial in view of their practical applications.

We present a detailed temperature dependent study of micro-photoluminescence (µPL) and single-photon emission to explore the physics of InGaAs/GaAs QDs grown by MOCVD with SRL and emitting at 1.3 µm. We identify excitonic complexes based on power-dependent and polarization-resolved µPL measurements und study them further by means of temperature-dependent µPL which allows us to identify the main µPL quenching mechanism in the investigated nanostructures. Furthermore, we probed single-photon emission with photon autocorrelation measurements in the temperature range of (5-50) K. The experimental findings allow us to point out the dominating factors limiting single-photon emission purity at elevated temperatures.

## Results

### Identification of excitonic complexes

In order to investigate in detail the temperature stability of emission and the impact of elevated temperatures on the single-photon purity from a single QD, excitonic complexes were first identified by means of excitation power-dependent and polarization-resolved µPL measurements. Owing to relatively small mesa size (disk shaped) of 1.3 µm and the QD spatial density of a few times $10^9/cm^2$ it was possible to spectrally isolate individual emission lines of high intensity and low background signal from a single QD at 5 K [see Fig. 1(a)]. Data from the same exemplary QD are presented throughout the manuscript. Different excitonic complexes, namely very prominent positive ($X^+$) trion, exciton ($X$), and for this particular QD also negative ($X^-$) trion as well as biexciton ($XX$) were pre-identified by means of excitation power-dependent µPL [Fig. 1(b)]. The observed line pattern is rather typical for this type of QDs[37]. We obtained linear scaling factors of emission intensity for exciton ($I_X \sim P^{1.08(5)}$), superlinear for trions ($I_{X^+} \sim P^{1.17(8)}$ and $I_{X^-} \sim P^{1.50(14)}$) and quadratic for biexciton ($I_{XX} \sim P^{1.95(22)}$), as expected[39,40]. Fig. 1(c) presents polarization-resolved µPL spectra, where we observed the fine structure splitting (FSS) of the neutral exciton. We determined FSS of $\Delta_{FSS} = 67$ µeV for this QD by fitting the polarization dependence of the emission energy with sine function [see Fig. 1(c) – right panel]. The extracted $\Delta_{FSS}$ value is within the range of FSSs typically observed for these QDs[37,41].

The trion complexes were identified based on the absence of FSS in polarization-resolved measurements supported by their superlinear power dependence. Noteworthy, a distinction between positively and negatively charged excitonic complexes was possible by comparing the emission spectrum with the results of 8-band k·p calculations of the QD's electronic structure combined with the configuration interaction method for excitonic states due to their significantly different binding energies[37]. Theoretical modelling (not shown here) reveals that these energies depend on exciton emission energy and for the reported QD, $X^-$ and $XX$ should be confined much stronger ($\Delta E_{X^-} \approx \Delta E_{XX} \approx 4$ meV) than $X^+$ ($\Delta E_{X^+} \approx 1$ meV).

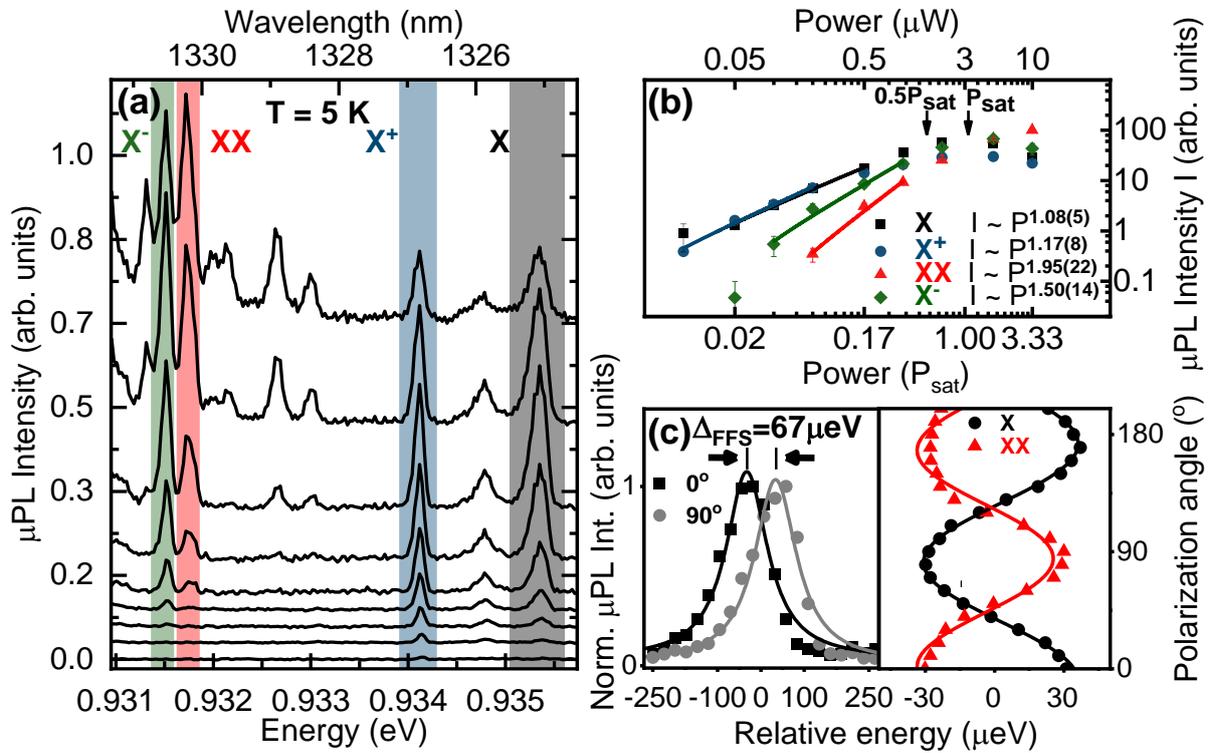

Fig. 1. µPL spectra from the investigated InGaAs/GaAs QD embedded in a mesa of 1.3 µmdiameter at 5 K. $X$ – exciton, $X^+$ – positive trion, $X^-$ – negative trion, $XX$ – biexciton. (a) Spectra for different excitation powers in the range of (0.05-15) µW. (b) Double logarithmic plot of the µPL intensity as a function of excitation power. Symbols denote the measured data points, the solid lines represent $I \propto P^\alpha$ dependence with a near quadratic (1.95) and linear (1.08) behavior for biexciton and exciton, respectively. Positive and negative trions exhibit superlinear dependence (1.17 and 1.50, respectively). (c) Polarization-resolved µPL spectra for perpendicular polarization angles for *X* in the left panel and polarization dependence of emission energy for *X* and *XX* in the right panel, both indicating exciton fine structure splitting of $\Delta_{FSS} = 67$ µeV.

## Temperature stability

To study the temperature stability of the QDs and to pinpoint the main carriers' escape channel from these QDs we performed µPL studies in the range of temperatures from 5 K to 80 K. The corresponding µPL spectra of the QD under study are presented in Fig. 2(a). Although the $X^+$ intensity was slightly lower than the *X* intensity at 5 K, it turned out to dominate the spectrum in the intermediate temperature range of 20 to 75 K. In fact, a pronounced $X^+$ µPL intensity increase was detected towards 30 K and the emission was still visible even above the liquid nitrogen temperature of 80 K. At this temperature the $X^+$ µPL intensity was reduced by a factor of 7 in comparison to its maximal value at 30 K.

A detailed quantitative analysis of the temperature dependence of emission from a studied QD is presented in Fig. 2(c). With increasing temperature, a redshift of all lines can be observed [see Fig. 2(a)-(b)] following the renormalisation of the QD material band gap energy ($E_g$). One can model the temperature dependence of energy gap taking into account temperature-dependent bosonic distribution of phonons by[42]:

$$E_g(T) = E_g(0) - S\langle\hbar\omega\rangle\left[\coth\left(\frac{\langle\hbar\omega\rangle}{2k_B T}\right) - 1\right]$$

with $E_g(0)$ the energy band gap at 0 K, *S* the electron-phonon coupling constant, $\langle\hbar\omega\rangle$ the average phonon energy, $k_B$ the Boltzmann constant, and *T* the temperature. Fitting the experimental data with the formula above allows to extract mean phonon energy of $\langle\hbar\omega\rangle$ = (8.0 ± 1.2) meV which corresponds well to the GaAs transversal acoustic phonons[43] confirming the predominant coupling of QD excitons with bulk phonons of the matrix material.

The data for the temperature dependence of linewidth *γ = γ(T)* [Fig. 2(d)] were fitted using a model including the linear contribution of acoustic phonons and exponential thermal activation of optical phonons with energy of $E_{LO}$, coupled to the exctionic complex[45]:

$$\gamma(T) = \gamma_0 + \gamma_{ac}T + \frac{\gamma_{LO}}{\exp(E_{LO}/k_B T) - 1}.$$

A fit to the temperature-induced linewidth broadening reveals at $T$ = 0 K line broadenings of approximately $\gamma_0$ = 80 µeV for $X^+$, $X^-$, $XX$ and $\gamma_0$ = 140 µeV for $X$. Here the difference in linewidths at low temperatures could be a result of stronger spectral diffusion (dynamical broadening due to charge fluctuations in the QD environment including defects in the SRL and charged surface states) caused by more pronounced Stark shift due to higher polarizability of the $X$ state than the other excitonic complexes[46,47].

The strong increase of linewidth setting in at about 40 K can be explained by the larger contribution of phonon sidebands to the total emission in this temperature range. Therefore, the determined full width at half maximum (FWHM) corresponds rather to the width of the phonon sidebands than to broadening of the zero-phonon line (ZPL)[48]. We found that the linewidths of the ZPL for all excitonic complexes are dominated by spectral diffusion: broadenings at the lowest investigated temperature are significantly greater than spectral resolution of our setup (< 25 µeV) and much larger than the lifetime-limited homogenous linewidth (0.7 µeV). We attribute the enhanced spectral diffusion to charged states at etched mesa surfaces. This interpretation is supported by the fact that even broader lines of (155 ± 50) µeV were observed in QDs embedded in smaller mesas with diameters of 0.3 µm. The higher optical quality of large mesas originates from the fact that in this case the probability that the QD is far away from the mesa sidewalls is higher and the influence of fluctuating charges trapped due to roughness of the etched mesa surface are a primary cause of spectral diffusion[44]. Increasing FWHM of emission is the reason for spectral lines to spectrally overlap and in consequence for the degradation of single-photon emission purity.

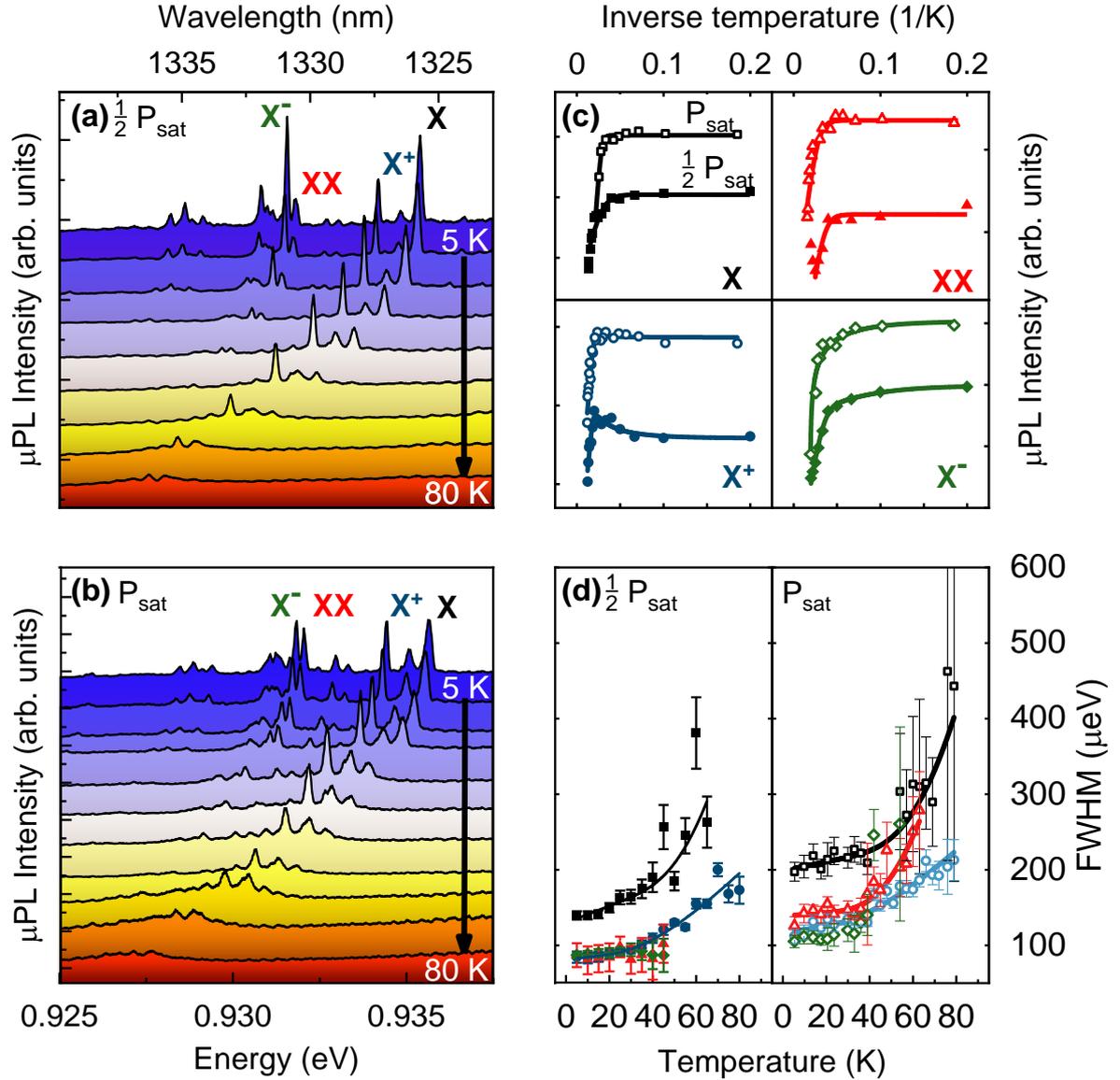

Fig. 2. (a)-(b) Temperature-driven evolution of μPL spectra from 5 K to 80 K for $\frac{1}{2}P_{sat}$ (a) and $P_{sat}$ (b). (c)-(d) Analysis of μPL temperature dependence for QD excitonic complexes: (c) μPL intensity (d) μPL linewidth. Symbols denote measured data, colored lines represent fits to the experimental data.

The changes of emission intensity shown in Fig. 2(a)-(c) are characteristic for the investigated type of QDs, similar results were obtained for other QDs in the same structure. To explain the μPL quenching [Fig. 2(c)], we compare our findings with recently reported results of 8-band k·p calculations for such dots which yielded the s-p splitting of (60-80) meV[38]. Here, the lower splitting corresponds to the higher QD emission energy. Noteworthy, the QDs analyzed here have even higher emission energies than those studied previously in Ref. 38. Additionally, 8-band k·p calculations predict s-p splitting to be shared by approximately (20-30) meV for holes and (40-50) meV for electrons. Analyzing the µPL temperature dependence we found characteristic activation energies responsible for quenching of individual emission lines to be in the range of (10-20) meV. This allows us to conclude that the dominant mechanism for the µPL quenching process is the promotion of holes to excited valence band states (confined levels) in the QD and subsequently their increasing thermal escape probability. The temperature-dependent data for *X* and *XX* are well explained by an Arrhenius dependence with a single activation energy. In contrast, for charged complexes a modification of the standard fitting formula was needed to account for the µPL intensity increase (decrease) in the low temperature range for *X$^+$* (*X$^-$*). As discussed in Ref. [36], this charging temperature effect can be related to thermal activation of one type of carriers from a reservoir, resulting in selective amplification (reduction) of *X$^+$* (*X$^-$*) emission at elevated temperatures [see Fig. 2(c)]. The temperature dependence of µPL intensity *I(T)* can be described by [35]:

$$I(T) = \frac{I_0 + I_P\left[1-\left(1-\frac{1}{1+B_1 exp(-E_1/k_B T)}\right)\right]}{1+B_2 exp(-E_2/k_B T)},$$

where *I$_0$*, *I$_P$* are initial and reservoir-gained intensities, *E$_1$*, *E$_2$* are activation energies of processes responsible for increase, decrease of emission intensity, respectively, and *B$_1$*, *B$_2$* are the amplitudes of thermally-activated processes with energies *E$_1$* and *E$_2$*. In the case of charged complexes, a distinct temperature interplay of their relative intensities was observed in the low temperature range: the activation energy *E$_1$* ≅ 1.9 meV was assigned to the pronounced µPL intensity increase of *X$^+$* at temperatures up to 30 K. This energy is characteristic also for *X$^-$* µPL intensity reduction indicating thermal activation of positive carrier traps (i.e. holes present due to unintentional carbon p-doping in MOCVD growth) in the vicinity of the QDs, resulting in higher probability of *X$^+$* formation. We could not observe clear correlations between the obtained activation energies for µPL thermal quenching (*E$_2$*) and the associated type of excitonic complex for the investigated QDs. This further indicates that the thermal intensity behavior is rather related to the s-p splitting and not to the QD size and specific type of excitonic complex. Then, the promotion of holes to higher QD states is the main path for thermal quenching of µPL intensity driven by the small interlevel spacing of the valence band confined states.

The temperature-dependent µPL study was carried out for two excitation regimes: at the exciton saturation power ($P_{sat}$ = 3.0 µW) and at a half of the saturation power (0.5 $P_{sat}$). The corresponding results are shown in Fig. 2(a)-(d). A qualitative difference for both excitation powers can only be observed in the case of the positively charged trion. Here, the activation energy of $E_1 \cong 1.9$ meV is visible only for the excitation power below saturation, for which the thermally activated carriers can still supply QDs and thus cause the $X^+$ µPL increase, and the simultaneously the $X^-$ µPL decreases. At $P_{sat}$, the trion µPL emission rate is limited by its radiative lifetime and even excess availability of thermally activated carriers cannot further increase the emission intensity. The observed relative increase in the intensity of $X^+$ with respect to $X$ can be related to increased probability of the QD being charged with residual carrier before the optical excitation.

Single-photon emission

Single-photon emission was probed by measuring the second-order photon autocorrelation function $g^{(2)}(\tau)$ and fitting the normalized histograms of coincidences with[49]:

$$g^{(2)}(\tau) = 1 - (1 - g^2(0)) \times e^{-\frac{|\tau|}{t_{rise}}},$$

where the function value for $\tau$ = 0 [$g^{(2)}(0)$] reflects the multiphoton emission probability, the antibunching time constant is $t_{rise} = \frac{1}{\Gamma + W_p}$, with the electron-hole radiative recombination rate $\Gamma$ and the effective pump rate $W_p$.

Experimental histograms are plotted together with the corresponding µPL spectra in Figs. 3(a)-(c). While $\Gamma$ (530 MHz for $X^+$, based on the time-resolved µPL experiment) is known to stay approximately constant with temperature[36], the observed decrease of $t_{rise}$ reflects the increase of the effective pump rate with temperature: the laser excitation power was increased with temperature to counteract the thermal quenching of µPL intensity. Moreover, $W_p$ increases also due to additional carriers, thermally activated from the reservoir (as discussed above for $X^+$). The values of multiphoton events' suppression and rise times obtained from fitting to the histograms of coincidences in Fig. 3(a)-(c) right panels are: (a) $g^{(2)}(0)$ = 0, $t_{rise}$ = 1.9 ns, (b) $g^{(2)}(0)$ = 0.17, $t_{rise}$ = 1.0 ns, and (c) $g^{(2)}(0)$ = 0.13, $t_{rise}$ = 0.64 ns with standard errors of fitted $g^{(2)}(0)$ values of 0.073, 0.11, 0.21, respectively. Here, the degradation of multiphoton suppression for elevated temperatures reflects the increasing contribution of uncorrelated background emission as seen in the associated µPL spectra due to temperature-induced quenching of QD emission. This contribution can be estimated by the ratio $\rho = S/(S + B)$[50] with intensity of signal $S$ and background $B$, which decreases from $\rho = 0.87$ at 5 K to $\rho = 0.57$ at 50 K. For all measurements the spectral filtering was kept constant and its bandwidth (0.43 nm) was in all cases broader than the FWHM of the $X^+$ emission line yielding 0.26 nm, 0.27 nm, and 0.40 nm for 5 K, 25 K, and 50 K, respectively. Therefore the low degradation of the $g^{(2)}(0)$ value is not achieved artificially by narrowing the spectral filtering, but is an inherent property of investigated structures.

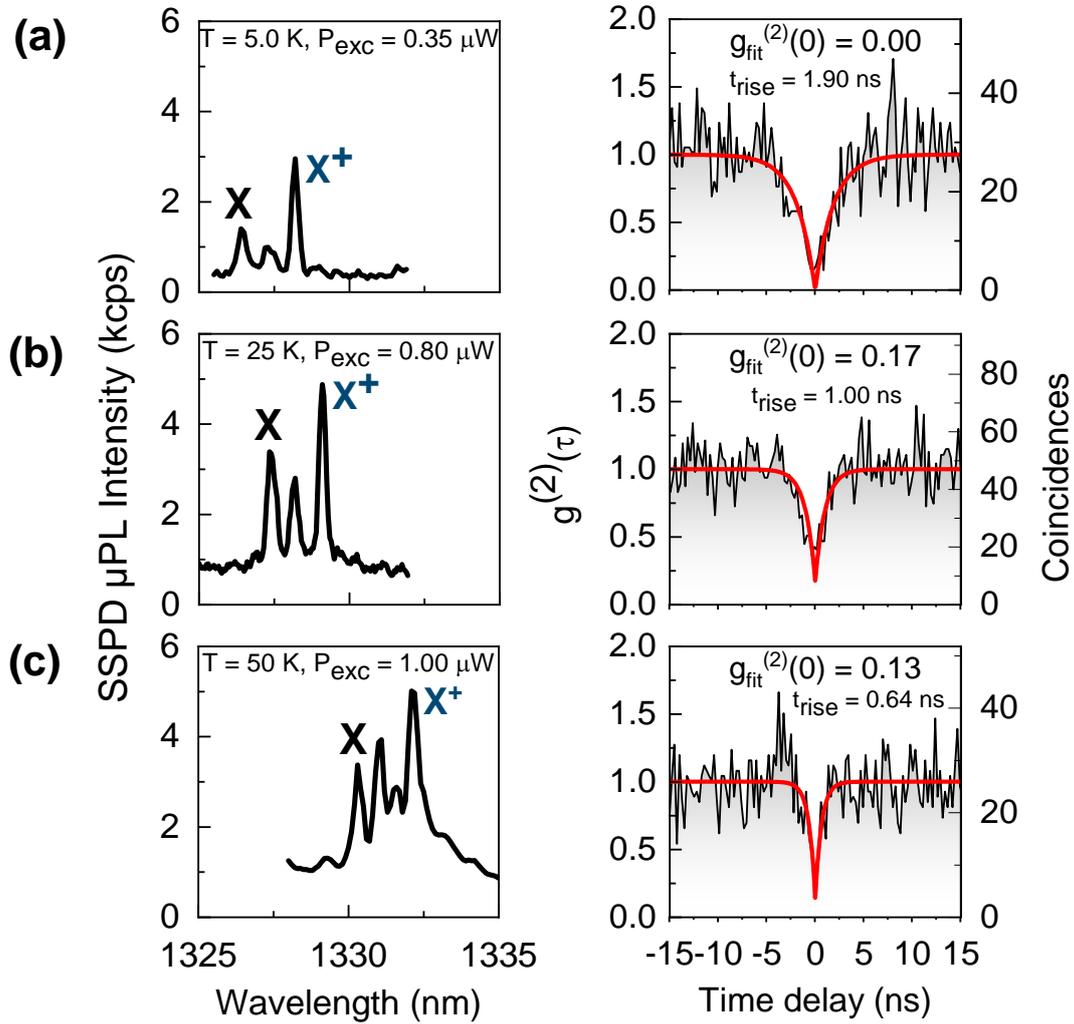

Fig. 3 Single-photon emission from the $X^+$ at cryogenic and elevated temperatures: spectra registered on superconducting single photon detectors - SSPD (left panel) and corresponding autocorrelation histograms (right panel), measured under continuous wave non-resonant excitation at (a) $T$ = 5 K, (b) $T$ = 25 K, and (c) $T$ = 50 K.

## Conclusions

In conclusion, we performed a detailed temperature- and power-dependent study of photoluminescence from single MOCVD-grown InGaAs/GaAs QDs, emitting in the telecom O-band at 1.3 µm, and embedded in deterministically fabricated mesas. Our studies provide important information about the temperature stability of single QD emission and assist at better understanding of processes involved in the temperature-induced degradation of single-photon emission purity. These processes include the increase of uncorrelated background photons and simultaneous decrease of the line intensity. Excitonic complexes in the QD were identified based on polarization-resolved and excitation-power-dependent µPL. The positive trion was found to be the thermally most stable excitonic complex, with its emission still visible at the temperature up to at least 80 K (liquid nitrogen accessible), and with single-photon emission purity of $g^{(2)}(0)$ = 0.13 at 50 K, what is particularly important in view of future applications due to applicability of cryogen-free cooling[23,24]. The positive trion emission intensity has a maximum around 30 K which is actually an ideal temperature when building a practical single-photon sources exploiting the Stirling cryocooling.

Our analysis indicates that the reason for non-monotonic temperature-dependence of the charged states is the thermal activation of holes from their traps in the QD vicinity. Moreover, we identified the excitation of holes to higher QD states in the valence band as the dominating mechanism for the µPL thermal quenching in the investigated range of temperatures. Both carrier loss and uncorrelated background photons at the same wavelength as the emitting line deteriorate the purity of single-photon generation from the studied QDs. While the uncorrelated background photons are a dominant process in multiphoton suppression degradation and could be limited by quasi-resonant[51,52] or resonant[53] excitation schemes, the reason for carrier loss, namely the holes' level separation which is responsible for the µPL quenching, would have to be tailored in the QD epitaxial growth (e. g. by including Al-containing barriers close to the QD/SLR to reduce thermal escape of carriers), in order to increase the thermal stability of the emission.

## Methods

Fabrication process

The investigated sample was grown by MOCVD on GaAs (001) substrate and consists of a 300 nm GaAs buffer layer, a distributed Bragg reflector composed of 23 pairs of GaAs/Al$_{0.9}$Ga$_{0.1}$As layers and a single layer of self-assembled In$_{0.75}$Ga$_{0.25}$As QDs grown in Stranski-Krastanow mode. Before the final 634 nm GaAs capping layer, a nominally 4 nm thick In$_{0.2}$Ga$_{0.8}$As SRL was deposited following the growth of QDs in order to redshift the emission down to 1.3 μm[17]. Mesas were deterministically fabricated by means of *in-situ* 3D electron-beam lithography[54,55]: first a bright QD was spectrally and spatially identified by means of low-temperature cathodoluminescence by applying electron dose low enough not to invert the CSAR62 resist spin-coated on the sample surface beforehand, and directly afterwards a cylindrical mesa coinciding in position with the pre-selected QD was defined in the resist by electron beam lithography. The sample was further dry etched using plasma reactive ion etching and the designed pattern was transferred from the resist to the GaAs capping layer. Mesas were fabricated with diameters in the range of 0.25 μm to 2.1 μm. All results presented in this manuscript were obtained for a QD in a mesa of 1.3 μm diameter.

Experimental setup

During spectroscopic measurements the structure was kept in the continuous-flow liquid-helium cryostat. Identification of excitonic complexes and temperature-dependent measurements were performed in a μPL setup equipped with a microscope objective with NA = 0.4, an 1 meter focal-length single-grating monochromator and InGaAs multichannel array detector (pixel size: 25 μm), providing spatial and spectral resolution of 2 μm and on the order of 25 μeV, respectively. Photon statistics measurements were conducted using a Hanbury Brown and Twiss setup configuration[56] with a monochromator of 0.32 m focal length acting as a band-pass filter with 0.43 nm bandwidth (kept constant for measurements at all temperatures). The filtered signal was coupled to a 50:50 fiber beam splitter. Each of its outputs was connected to NbN superconducting single photon counting detectors with ~10-15% quantum efficiency and 10 dark counts/s at 1.3 μm. The photon correlation statistics was acquired by a multichannel picosecond event timer with time-bin width set to 256 ps. In both types of experiments the QD was excited non-resonantly by a continuous wave semiconductor diode laser with energy of 1.88 eV (λ = 660 nm) and 1.56 eV (λ = 787 nm) for the temperature-dependent μPL and autocorrelation measurements, respectively.

## Acknowledgements

The authors acknowledge financial support via the "Quantum dot-based indistinguishable and entangled photon sources at telecom wavelengths" project, carried out within the HOMING programme of the Foundation for Polish Science co-financed by the EU under the EFRE. Support from the Polish National Agency for Academic Exchange is also acknowledged. This work was also funded by the FI-SEQUR project jointly financed by the European Regional Development Fund of the European Union in the framework of the programme to promote research, innovation, and technologies (Pro FIT) in Germany, and the National Centre for Research and Development in Poland within the 2nd Poland-Berlin Photonics Programme, grant number 2/POLBER-2/2016 (project value 2 089 498 PLN). Support from the German Science Foundation via CRC 787 is also acknowledged.

## Author contributions

P. H. and A. M. performed autocorrelation measurements. P. H. and M. B., supervised by A.M., performed micro-photoluminescence measurements and wrote the first version of the manuscript. D. Q. designed and grew the sample under supervision of A. S.; N. S. patterned the sample under supervision of S. Rodt. S. Reitzenstein and G. S. supervised the work of the Berlin and Wroclaw teams, respectively. All authors took part in the preparation of the final version of the paper. P. H. and M. B. contributed equally to this work.

## Competing interests

The author(s) declare no competing interests.